\documentclass[prb,twocolumn,showpacs,superscriptaddress]{revtex4}
\usepackage{graphicx,dcolumn,amsmath}

\begin{document}

\title{\textbf{ Optical properties of structurally-relaxed Si/SiO$_2$
superlattices:  the role of bonding at interfaces. }}

\author{\firstname{Pierre} \surname{Carrier}} \affiliation{D\'{e}partement
de Physique et Groupe de Recherche en Physique \\et Technologie des
Couches Minces (GCM), Universit\'{e} de Montr\'{e}al, \\ Case Postale
6128, Succursale~Centre-Ville, Montr\'{e}al, Qu\'{e}bec, Canada H3C 3J7}

\author{\firstname{Laurent J.} \surname{Lewis}} \email[Author to whom
correspondence should be addressed: Email address:\ ]
{laurent.lewis@umontreal.ca} \affiliation{D\'{e}partement de Physique et
Groupe de Recherche en Physique \\et Technologie des Couches Minces (GCM),
Universit\'{e} de Montr\'{e}al, \\ Case Postale 6128,
Succursale~Centre-Ville, Montr\'{e}al, Qu\'{e}bec, Canada H3C 3J7}

\author{\firstname{M. W. C.} \surname{Dharma-wardana}}
\affiliation{Institute for Microstructural Sciences, National Research
Council, Ottawa, Canada K1A 0R6}

\date{\today}

\begin{abstract}

We have constructed microscopic, structurally-relaxed atomistic models of
Si/SiO$_2$ superlattices. The structural distortion and oxidation-state
characteristics of the interface Si atoms are examined in detail. The role
played by the interface Si suboxides in raising the band gap and producing
dispersionless energy bands is established. The suboxide atoms are shown
to generate an abrupt interface layer about 1.60 \AA\ thick. Bandstructure
and optical-absorption calculations at the Fermi Golden rule level are
used to demonstrate that increasing confinement leads to (a) direct
bandgaps, (b) a blue shift in the spectrum, and (c) an enhancement of the
absorption intensity in the threshold-energy region. Some aspects of this
behaviour appear not only in the symmetry direction associated with the
superlattice axis, but also in the orthogonal plane directions. We
conclude that, in contrast to Si/Ge, Si/SiO$_2$ superlattices show clear
optical enhancement and a shift of the optical spectrum into the region
useful for many opto-electronic applications.

\end{abstract}


\maketitle

\section{Introduction}\label{Intro}

The initial interest in light-emitting Si-based nanostructures has lead to
a number of important experiments establishing that Si/SiO$_2$
superlattices (SLs) show enhanced, blue-shifted luminescence.  
\cite{LuLockBar,Kanemitsu,Novikov,Khriachtchev,Mulloni,Kanemitsu_bis}
While the luminescence pattern is more complex in these systems than in
others, the blue shift was found to correlate with decreasing Si-layer
thickness. This simple relation between the silicon-layer thickness and
the luminescence peak is of great interest for applications such as
Si-based light-emitting diodes (\mbox{Si-LEDs}). All reported SL energy
peaks are in the lower part of the visible spectrum --- the highest
reported value being \mbox{2.3 eV} (540 nm), i.e., green,\cite{LuLockBar}
and the lowest \mbox{1.2 eV} (1030 nm), in the near
infrared.\cite{Novikov} These SLs are really multiple Si quantum wells
(MQW), the silicon oxide layers playing the role of barriers. The
thickness of the Si quantum wells, $L_{\rm Si}$, is the critical
parameter. A fixed $L_{\rm Si}$ would simply set the colour of the
\mbox{Si-LED}, while MQWs with a range of $L_{\rm Si}$ would span a range
of colours in the luminescence.

Other systems containing confined silicon structures (besides MQWs) are,
e.g., porous Si\cite{Canham,LehmanGosele} (consisting of quasi
one-dimensional structures), silicon nanoclusters in SiO$_2$
matrices,\cite{Tsybeskov} nanocrystals\cite{Cheylan} or dislocation loops,
and quantum-dot structures made from implantation of boron\cite{Ng} or
other ions.\cite{ericson} The choice of a particular structure, e.g., for
\mbox{Si-LED} applications, depends on many factors: stability over time,
optical efficiency at room temperature, experimental reproductiveness,
facility to accept \mbox{n-} or \mbox{p-} type dopants (e.g., for n-p
junctions), ease of incorporation in ultra-large-scale-integration
technology, and production costs. \mbox{Si/SiO$_2$} SLs are stable
structures, as opposed to porous-Si. In addition, the silicon layer
thicknesses in SLs are directly related to the energy peaks in the
Si/SiO$_2$ luminescence spectra. This is not straighforward in porous
silicon, Si clusters or nanocrystals, where pore dimensions as well as
hydrogen concentrations play an uncontrolled role on the energy
shift.\cite{Tsybeskov,porous}

Silicon-based structures have many advantages over structures made of
other semiconductors: low-cost (as compared to any of the \mbox{III-V's}
or \mbox{II-VI's}), non-toxicity, practically unlimited availability (in
contrast to germanium) and benefits from decades of experience in
purification, growth, and device fabrication. However, the indirect energy
gap (\mbox{$\sim$1.1 eV}) in bulk crystalline silicon (\mbox{c-Si}) makes
it unsuitable for optoelectronic applications. Silica (SiO$_2$) is another
key material of the microelectronics industry; it has a bandgap of
\mbox{$\sim$9 eV}. Optical-fiber technologies and
metal-oxide-semiconductor field-effect transistors (MOSFETs) are based on
(high-quality) silica. Molecular-beam epitaxy and chemical vapor
deposition provide the needed growth technology for Si and SiO$_2$. It is
possible to combine crystalline
silicon\cite{Kanemitsu,Luprivate,Zacharias,ZachariasAPL} and SiO$_2$ to
produce structured materials having chemically pure, sharp, defect-free
interfaces.  The degree of advancement of the fabrication technology is
such that the enhanced luminescence in \mbox{Si/SiO$_2$} SLs cannot be
explained only in terms of defects\cite{Kanemitsu_bis} and/or residual
hydrogen atoms filling unsaturated Si dangling bonds --- the latter being
referred to as $P_b$-type centers\cite{Stirling} --- located at the SL
interfaces, as in the well-understood enhanced luminescence of
hydrogenated {\em a-}Si.

The objective of the present article is to consider the detailed
microscopic structure of the \mbox{SiO$_2$/Si/SiO$_2$} double interface
structure and provide a first-principles understanding of the emission of
light from \mbox{Si/SiO$_2$} SLs. Only a few atomistic DFT calculations on
\mbox{Si/SiO$_2$} SLs have been
reported.\cite{Delley,Kageshima,Pukkinen,Degoli,Agrawal,Carrier} Many more
would have been available were it not of the (naturally-occuring)
amorphous structure of SiO$_2$: amorphous structures require large
supercells to get physically relevant results. Another important issue is
the need to model the Si/SiO$_2$ interface so as to correctly incorporate
the known experimental details. Recent core-level shift
measurements\cite{Fuggle} provide details of the suboxide (partially
oxidized) Si atoms.\cite{SiegerLuHimpsel} Further, the abruptness of the
Si/SiO$_2$ interface has been established from transmission electronic
microscopy (TEM) experiments; values of the interface width as low as 5
\AA\ have been reported.\cite{Lockwood,Kanemitsu} Based on these
observations, realistic Si/SiO$_2$ interface models have been designed by
several
workers.\cite{Pasquarello,Neaton,Tu,NgInterface,Stirling,KageshimaInterface}
Suboxide Si atoms in most of these models are distributed within three
atomic layers of the Si/SiO$_2$ interface, corresponding to the lowest
experimental interface thickness. Such interface models are needed for
first-principles modelling of MOSFETs, and for SL structures, where
\textit{multiple} Si/SiO$_2$ interfaces are present.

An early (and tractable) Si/SiO$_2$ interface model was proposed by Herman
and Batra.\cite{HermanBatra} The large lattice mismatch between SiO$_2$
and Si was accommodated by setting the \mbox{$\beta$-cristobalite} SiO$_2$
unit cell diagonally on the diamond-like \mbox{c-Si} unit cell. An oxygen
atom was included in the interface to saturate the dangling bonds,
resulting in a crystalline model of the interface. Another simple
crystalline model, involving a bridge oxygen at the interface, was
introduced by Tit and Dharma-wardana.\cite{TitDharmaDBMBOM} These models
were studied using a variety of methods --- tight-binding
(TB),\cite{HermanBatra,TitDharmaDBMBOM} full-potential,
linear-muffin-tin-orbitals (FP-LMTO)\cite{Pukkinen,Degoli} and
linearized-augmented-plane-waves (FP-LAPW).\cite{Carrier} The
dispersionless character of the bandstructure in the growth axis has been
confirmed within all three theoretical approaches, thus demonstrating the
existence of strong confinement. However, the nature of the energy gap ---
from both LMTO and LAPW calculations --- is still indirect. Furthermore,
these crystalline models show that the light absorption is quite dependent
on the details of bonding and interface structure (bondlengths, angles and
chemical species), emphasizing the need for \emph{more realistic},
\emph{structurally-relaxed} models.\cite{Carrier}

Several structurally-relaxed models have been investigated by Kageshima
and Shiraishi using first-principles methods.  Their models were
constructed starting from $\beta$-cristobalite as well as $\alpha$-quartz
SiO$_2$ layers superposed onto c-Si layers with different possibilities
for the Si/SiO$_2$ interface (such as hydrogen atoms at dangling
bonds);\cite{Kageshima} the atomic sites were then structurally relaxed.
The calculations indicate that the energy gaps are indeed direct, and that
interfacial Si--OH bonds are possible candidates for the light-emitting
enhancement in SLs.  However, the models are not consistent with the
observed suboxide Si atomic distributions at the Si(001)/SiO$_2$
interfaces. For instance, no Si atoms bonded to three oxygens are present.
Tit and Dharma-wardana\cite{TitDharma} have constructed a
partially-relaxed model (PRM) starting from a structurally relaxed
Si(001)/SiO$_2$ interface structure due to Pasquarello, Hybertsen, and Car
(PHC).\cite{Pasquarello} This interface model contains all three suboxide
Si atomic species. Hydrogen atoms were used by PHC to terminate the
surface. In the Tit and Dharma-wardana model, the H atoms were removed and
the Si/SiO$_2$ interface structure was converted into a SiO$_2$/Si/SiO$_2$
double interface SL structure while preserving local tetrahedral bonding
to obtain the PRM structure. Within the TB approach, the energy gap of the
PRM was shown to be direct and enhancement of the optical absorption (as
compared to c-Si) was confirmed.\cite{tran} This is further described in
the next section, since the fully relaxed models (FRMs) discussed in the
present work are based on this PRM.

The results presented here go beyond the TB approach and were obtained
within the projector-augmented wave (PAW) theoretical framework. A brief
description of the theory is given in section \ref{CompDet}. The SL models
were all structurally relaxed, and contain no hydrogen atoms at the
interfaces. The interface suboxide Si atoms observed in
experiment\cite{SiegerLuHimpsel} arise naturally in the model.
Calculations were carried out for different Si-layer thicknesses, in order
to assess the effect of confinement on the electronic and optical
properties. Supplementary models have been constructed to clarify the role
of the suboxide atoms on the SL optical properties. We first review the
theoretical methods (section \ref{CompDet}), then focus on the four
central issues needed to understand the Si/SiO$_2$ luminescence
properties: construction of realistic interface models
(section~\ref{Ionic}), quantum confinement (section~\ref{confinement}),
role of the suboxide atoms (section~\ref{interface}), and optical effects
associated with increased confinement (section~\ref{blueshift}). Our study
thus provides a complete, microscopic picture of the luminescence
properties of Si/SiO$_2$ SLs.

\section{Computational details}\label{CompDet}

 The electronic-structure calculations were carried out using the Vienna
\textit{ab initio} simulation package ({\sf VASP})\cite{VASPref} using the
``frozen-core'' PAW approach.\cite{KresseJoubert} The overall framework is
density-functional theory\cite{HohenbergKohn,KohnSham} (DFT) within the
local-density approximation (LDA).\cite{Payne}

The ``frozen-core'' PAW is a simpler form of the general PAW method
introduced by Bl\"ochl.\cite{BlochlPAW} Bl\"ochl's method is an extension
of the usual LAPW\cite{Singh} approach. Hence, the PAW method formally
bridges the LAPW to the ultrasoft-pseudopotentials (US-PP) in order to
combine the precision of the former and the rapidity (for larger systems)
of the latter. The PAW method has another advantage over the usual
implementation of US-PP, essential for optical calculations: it avoids
correcting for spatial nonlocality effects in typical pseudopotentials
when evaluating the momentum operator \textbf{p}. Further details can be
found in the article of Adolph \emph{et al}.\cite{Adolph}

Matrix elements of the momentum operator \textbf{p} are needed for
calculating interband optical effects. We describe the main steps that
lead to the calculation of the absorption coefficient, starting from PAW
solutions of the Kohn-Sham equations. The PAW approach rests on the
following linear transformation:
  $$
  \vert{\Psi}_N\rangle = \vert\tilde{\Psi}_N\rangle +
  \sum_i \left(\vert{\phi}_N\rangle-\vert\tilde{\phi}_N\rangle\right)
  \langle\tilde{p}_i\vert\tilde{\Psi}_N\rangle.
  $$
  This relates the (calculated) pseudo wave function
$\vert\tilde{\Psi}_N\rangle$ to the (now corrected) all-electrons wave
function $\vert{\Psi}_N\rangle$. The index specifies the atomic site,
angular momentum numbers and reference energy. The two functions
$\vert\tilde{\phi}_N\rangle$ and $\vert{\phi}_N\rangle$ are respectively
the pseudo, and all-electron wave function of a reference atom. They are
forced to overlap outside a given core region. The functions
$\vert\tilde{p}_i\rangle$ are the \textit{projector} functions
characteristic of the PAW method. Thus, the three functions
$\vert{\phi}_N\rangle$, $\vert\tilde{\phi}_N\rangle$ and
$\vert\tilde{p}_i\rangle$ constitute the frozen-core PAW data, being set
prior to self-consistent field calculations. The projector functions
$\vert\tilde{p}_i\rangle$ are constructed so as to remain dual to the
pseudo wave functions, to fulfill generalized orthogonality constraints
and to remain (approximately) complete (see Ref.~\onlinecite{Adolph} for
full definitions).

The application of the above linear transformation to any operator $A$
within the PAW approach has been described by Bl\"ochl [cf.\ eq.\ (11) of
Ref.~\onlinecite{BlochlPAW}]. If $A$ is the momentum operator \textbf{p}
(in a certain direction defined by the polarization vector), we have
  \begin{eqnarray*}
  \lefteqn{
  \langle{\Psi}_N\vert\mbox{{\textbf{p}}}\vert{\Psi}_M\rangle =
  \langle\tilde{\Psi}_N\vert
  \mbox{{\textbf{p}}}\vert\tilde{\Psi}_M\rangle
  + } \\ & & \sum_{i,j} \langle
  \tilde{\Psi}_N\vert\tilde{p}_i\rangle\left(
  \langle{\phi}_i\vert\mbox{{\textbf{p}}}\vert{\phi}_j\rangle -
  \langle\tilde{\phi}_i\vert
  \mbox{{\textbf{p}}}\vert\tilde{\phi}_j\rangle
  \right) \langle \tilde{p}_j\vert\tilde{\Psi}_M\rangle.
  \end{eqnarray*}
  The imaginary part of the dielectric function, $\epsilon_i$, can be
determined by using the Fermi Golden rule within the Coulomb gauge; the
expression becomes:\cite{Cardona}
  \begin{eqnarray*}
  \displaystyle
  \epsilon_i(E) = & \frac{\kappa^2}{E^2}
  \displaystyle
  \sum_{M,N} \int_{BZ} \frac{2d^3\vec{k}}{(2\pi)^3}
  \displaystyle
  \vert\langle{\Psi}_N\vert
  \mbox{{\textbf{p}}}\vert{\Psi}_M\rangle\vert^2
  \\ \nonumber
  & \!\!\times f_{N}(1\!\!-\!\!f_{M})
  \delta(E_{N} - E_{M} -
  E), \\ \nonumber
  \end{eqnarray*}
  where $\kappa=2\pi e/m$. The function $f_{n}$ is the Fermi distribution
and $\langle{\Psi}_N\vert\mbox{{\textbf{p}}}\vert{\Psi}_M\rangle$ are the
PAW matrix elements. The whole expression corresponds to the probability
per unit volume for a transition of an electron from the valence band
state $\vert{\Psi}_N\rangle$ to the conduction band state
$\vert{\Psi}_M\rangle$ to occur.

The tetrahedron method\cite{BlochlTET} is used to evaluate
$\epsilon_i(E)$.  The joint density of states, which determines the
interband transitions $\delta(E_{N} - E_{M} - E)$ and the optical matrix
elements
$\vert\langle{\Psi}_N\vert\mbox{{\textbf{p}}}\vert{\Psi}_M\rangle\vert^2$,
are computed on each tetrahedron (i.e., 1/4 $\times$ the sum of the matrix
elements on the four corners of each tetrahedron). The real part
$\epsilon_r$ is then obtained using the Kramers-Kronig
relation.\cite{Cardona} Since the dielectric function is the square of the
complex refractive index, $(\epsilon_r + i \epsilon_i) = (n_r + i n_i)^2$,
the absorption coefficient becomes
   $$
   \alpha(E)= 4\pi \frac{E}{hc} n_i =
   4\pi \frac{E}{hc} \left[\frac{(\epsilon_r^2 + \epsilon_i^2)^{1/2} -
   \epsilon_r}{2}\right]^{1/2}
   $$
 with $c$ the speed of light in vacuum, $h$ Planck's constant, and $E$ the
photon energy.

Electron-hole (e-h, excitonic) interactions were not included in the
calculations, as they would be in, say, solutions of the Bethe-Salpeter
equation. The size of our systems prohibits such complete optical
calculations, which are feasible only for very small systems (a few
atoms). These additional effects would enhance the results from interband
transitions since e-h interactions generally increase the absorption at
the onset.\cite{Chang}

\section{Construction of the structural models}\label{Ionic}

Recent core-level shift experiments\cite{SiegerLuHimpsel} have revealed
the presence of all possible oxidation states for Si atoms in Si/SiO$_2$
structures such as SLs, that is Si$^{+n}$, where $n= 0,1,2,3,4$ is the
charge found within each Si Wigner-Seitz sphere. Si$^{0}$ and Si$^{+4}$
are the charge states of Si found in bulk-Si and bulk-SiO$_2$. The
suboxide (subO) Si atoms with $n=1,2,3$ are found at the interface.
Slightly larger distributions for the subO Si$^{+3}$ densities, as
compared to those for Si$^{+1}$ and Si$^{+2}$, were reported in these
experiments. Microscopic Si/SiO$_2$ interface models should be consistent
with experiments in closely reproducing the density distributions of
\emph{all} subO Si atoms.

As mentioned above, the Si/SiO$_2$ SL model structures discussed in the
present article are based on the Si(001)/SiO$_2$ interface structures
obtained by PHC,\cite{Pasquarello} who used the Car-Parrinello method to
relax the models to their energy minima. It is important to note that the
PHC models were \emph{not} designed for the double interface structure
found in SiO$_2$/Si/SiO$_2$ SLs but, rather, for a single Si/SiO$_2$
interface which terminates into the vacuum; this is done by saturating
dangling bonds with H atoms. Thus, these models {\it de facto} contain the
essential details of atomic positions and charge states at the Si/SiO$_2$
interface.

\begin{figure}[htb]
   \includegraphics*[width=6cm]{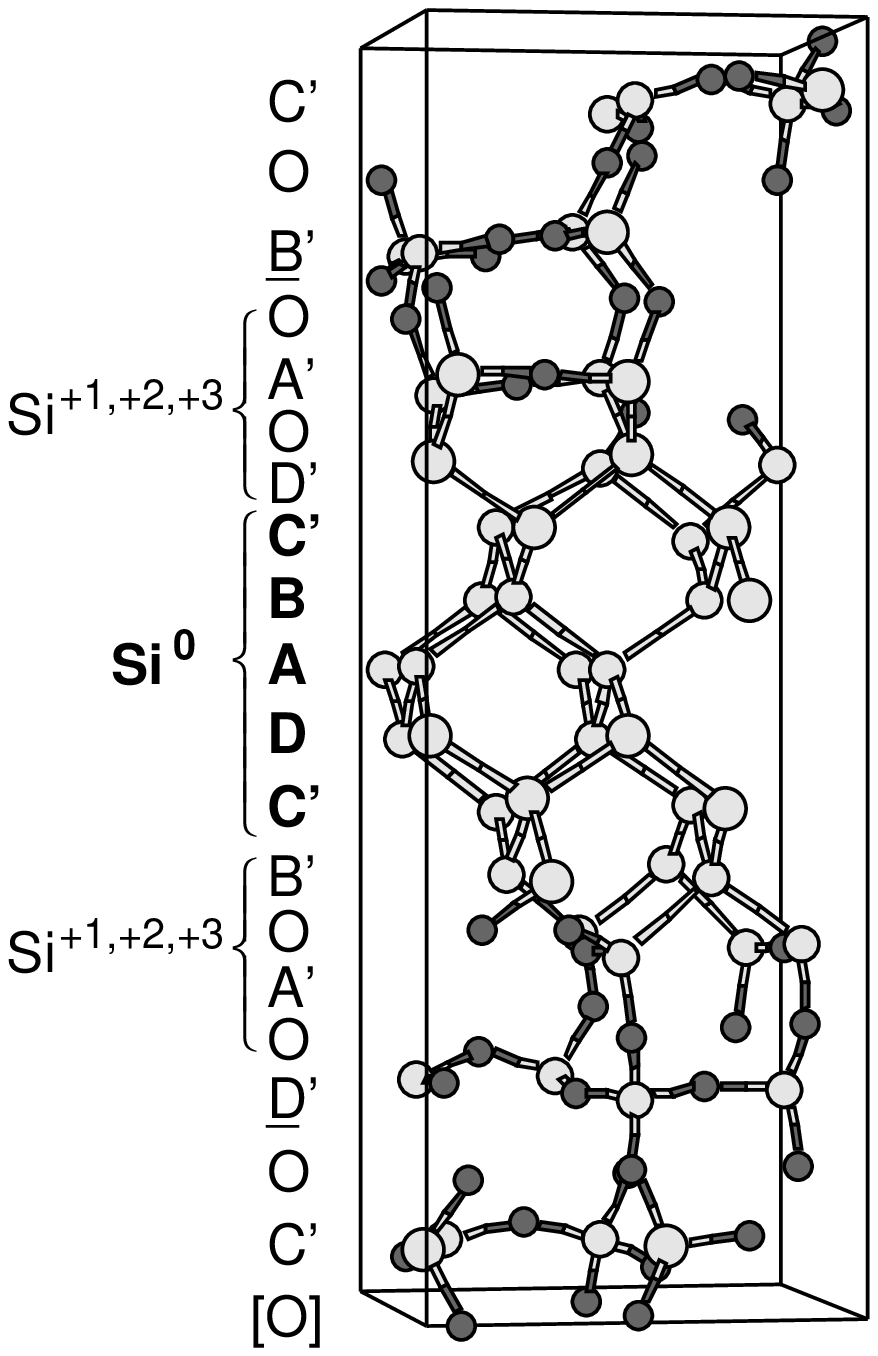}
  \caption{ The unit cell of the fully relaxed SL model (FRM1).
  The configuration of
  the bulk-like and suboxide Si atomic planes is also depicted.
  The white and black circles are respectively the positions of
  Si and O atoms. }
  \label{modelSL}
  \end{figure}

Tit and Dharma-wardana\cite{TitDharma} have generated a Si/SiO$_2$ SL
structure starting from one of the PHC models that contains an
\emph{equal} distribution of the three subO atoms, in (almost complete)
accord with experiment. This SL model has been constructed by first
operating a mirror transformation and then a partial rotation of the
Si/SiO$_2$ section of the relaxed interface structure, leading to an
intermediate Si/SiO$_2[\mbox{mirror}]$SiO$_2$/Si SL structure. Second,
some of the Si layers were inverted in order to meet the $sp^3$-bonding
topology. The resulting Si/SiO$_2$ SLs structure, fully described in the
article of Tit and Dharma-wardana,\cite{TitDharma} has the following final
configuration,
  $$
  \overbrace{[O]C^{'}O\underline{D}^{'}O}^{\mbox{Si$^{+4}$}}
  \underbrace{A^{'}OB^{'}}_{\mbox{Si}^{+1,2,3}}
  \underbrace{\bf C^{'}\stackrel{\Downarrow}{\displaystyle
  \bf DA}\bf BC^{'}}_{\mbox{Si$^0$}}
  \underbrace{D^{'}OA^{'}}_{\mbox{Si}^{+1,2,3}}
  \overbrace{O\underline{B}^{'}OC^{'}}^{\mbox{Si$^{+4}$}}.
  $$
 The connection between the symbols in this configuration and the specific
atomic layers in the model is shown in Fig.~\ref{modelSL}. The letters
$A$, $B$, $C$ and $D$ correspond to silicon atomic layers, while the $O$s
are oxygen layers. The primes denote layers that depart from the
diamond-like-Si crystalline arrangement, and $[O]$ corresponds to the
layer where the mirror operation has been performed. $\underline{B}^{'}$
and $\underline{D}^{'}$ are the Si layers which have been inverted in
order to satisfy the $sp^3$-bonding topology. The subO Si$^{+n}$ atoms,
with $n=1,2,3$, are distributed within only two Si layers, while the
Si$^{0}$ atoms are distributed within five atomic layers. The embryonic
PHC interface model corresponds roughly to one side of the above
configuration, starting from the arrow up to the right.

Of course, this construction induces artificial symmetries in the middle
of the SiO$_2$ layer (more precisely, upon and around the $[O]$ layer);
this model is thus in essence \emph{partially} relaxed --- the PRM
referred to earlier. Significant information on the electronic and optical
properties of this model have been extracted, within the TB approach, by
Tit and Dharma-wardana,\cite{TitDharma} who obtained direct energy gaps as
well as dispersionless bandstructures. Furthermore, the imaginary part of
the dielectric function was calculated, and then the absorption
coefficient was deduced. From this calculation, enhancement of absorption
as well as blueshift with confinement have been demonstrated.\cite{tran}

\begin{figure}[htb]
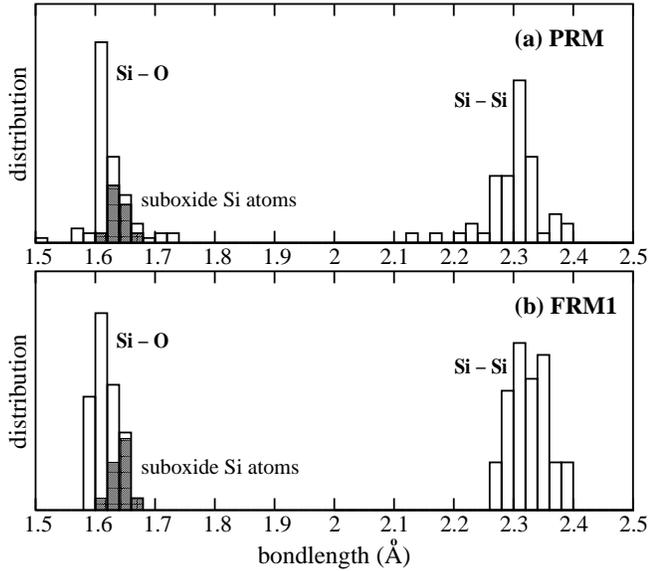

   \includegraphics*[width=8.5cm]{distPRM.eps}
   \includegraphics*[width=8.5cm]{distFRM1.eps}
  \caption{ Evolution of the Si--O and Si--Si bondlengths from (a) the PRM
  construction by Tit and Dharma-wardana, to (b)
  the fully relaxed structure (FRM1) described in the text. }
  \label{PRM2FRM}
  \end{figure}

The next obvious step is to relax the PRM, i.e., determine the set of
positions which leads, via the Hellman-Feynman forces, to the lowest total
energy. We have used the PAW approach described in the previous section to
obtain a first fully-relaxed model (FRM1), which contains approximately
one unit cell of confined-Si. The supercell contains 52 Si and 44 O atoms,
and has dimensions \mbox{7.675 $\times$ 7.675 $\times$ 24.621 \AA$^3$}.  
The relaxation procedure has been performed with five \textbf{k} points in
the reduced Brillouin zone (BZ). The energy cutoff was 25.96 Ryd in all
calculations. The total energy was found to decrease by 30.84 eV (0.32 eV
per atom) during relaxation.  Figure \ref{PRM2FRM} shows the bondlength
distributions before (PRM) and after (FRM1) relaxation; the bondlengths
are centered around the expected values, viz., $\sim$ 1.61 \AA\ for Si--O
and 2.35 \AA\ for Si--Si bonds. The shaded boxes in Fig.~\ref{PRM2FRM} are
the distributions of the interface subO Si atoms, while the empty boxes
are the total distribution bondlengths, including subO Si atoms; the
shaded boxes remain relatively unchanged upon relaxation since both
interfaces were already at their energy minimum, after PHC. The main
atomic drift during relaxation occurs in the center of the Si and SiO$_2$
layers, i.e., near the [O] and the $A$ layer of the configuration
discussed above. Interfacial Si--O bondlengths of all subO Si atoms depart
from the values of Si$^{+4}$ in the SiO$_2$ layer. The broadening of the
Si--Si bondlengths is in general much larger than that of Si--O; the
distortion of the bondlengths are thus mainly within the Si layer and at
the Si/SiO$_2$ interface, i.e., not inside the silica layer.  This is
further discussed below. The resulting FRM1 is shown in
Fig.~\ref{modelSL}.

Additional models having thicker Si wells were constructed in order to
examine the role of subO Si layers and the effect of confinement on the
electronic and optical properties. As noted earlier, the FRM1 structure
contains approximately one unit cell of confined-Si, i.e., the set of
layers with charge state Si$^0$ (bulk-like Si).  By inserting one, then
two, additional $ABCD$ Si atomic planes (i.e., one Si unit cell, thickness
5.43 \AA), and relaxing all atoms, we generated two additional models --
FRM2 and FRM3. The FRM2 contains 68 Si atoms while the FRM3 has 84 Si
atoms; both have 44 oxygen atoms, as in the FRM1. The {\em total} energy
variation for the FRM2 during relaxation was found to be only 0.15 eV
(i.e., 0.0013 eV/atom) while for the FRM3, this change is a minuscule
0.051 eV (i.e., 0.00040 eV/atom). These numbers imply that both FRM2 and
FRM3 have essentially crystalline Si layers. The FRM2 has nine Si$^{0}$
atomic planes while the FRM3 has thirteen.

\begin{figure}[htb]
   \includegraphics*[width=8.5cm]{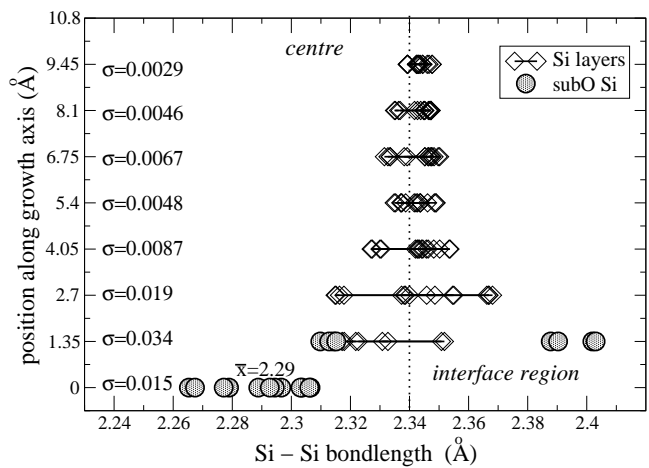}
  \caption{ Si--Si bondlength distribution in the FRM3, from the interface
  (bottom of figure)
  towards the center (top of figure) of the Si layer. (There are thirteen
  Si$^0$ layers in the FRM3 and thus six interplanar Si--Si bondlengths
  starting from both interfaces towards the center of the Si layer.}
   \label{couches}
  \end{figure}

Figure \ref{couches} shows the distributions of the Si--Si bondlengths in
the FRM3, starting from the Si(001)/SiO$_2$ interfaces (at the bottom of
Fig.~\ref{couches})  and going towards the centre of the silicon layer
along the growth axis.  The standard deviation ($\sigma$) from the mean
value ($\bar{x}$=2.34 \AA\ in all atomic layers except at the interface,
where $\bar{x}$=2.29\AA\ ) is also given.  The diamond-shaped symbols
correspond to the Si--Si bondlengths for Si$^0$ atoms while the filled
circles are subO Si--Si bondlengths at the interfaces. The Si--Si
bondlengths depart significantly from their crystalline counterparts at
the interfaces up to about three atomic layers, where $\sigma$=0.019; the
standard deviation is four times higher at the interfaces than in the
fifth atomic layer. This deviation of the bondlengths at the Si/SiO$_2$
interfaces shows that it is important to take relaxation aspects into
account.  Indeed, amorphous SiO$_2$ layers in realistic SLs induce strain
and disorder in the Si layer, as confirmed in Fig.~\ref{couches}. This
effect could generate localized defects giving rise to efficient radiative
electron-hole recombination. However, the strain fields only contribute to
the small \emph{quasi-momenta} regime and cannot easily supply the
momentum deficit involved in the indirect transition of c-Si. Moreover,
the relatively small size of our supercell models in the $x$-$y$
directions prevents firm conclusions being drawn about the influence of
this strain on the optical properties. Further aspects of the role of
interfaces are discussed in section \ref{interface}.

\begin{figure}[htb]
   \includegraphics*[width=6cm]{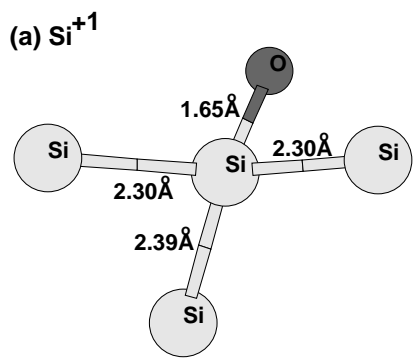}
   \includegraphics*[width=6cm]{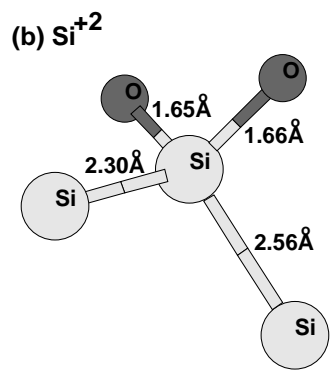}
   \includegraphics*[width=6cm]{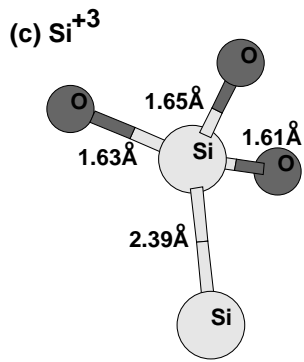}
  \caption{Structure of the three suboxide interfacial
           Si atoms.}
  \label{tetra}
  \end{figure}

The three subO Si configurations at the interfaces of the FRMs are shown
in Fig.~\ref{tetra} with their corresponding bondlengths. Note that the
left and right interfaces in the SiO$_2$/Si/SiO$_2$ SLs are not
\emph{exactly} equivalent; they remain independent (during structural
relaxation, for instance). However, a subO Si on the left interface has a
locally equivalent subO Si on the right interface, by construction. As a
consequence, each pair of equivalent subO Si atoms have approximately the
same bondlengths and angles in all the FRMs. As seen in Fig.~\ref{tetra},
the bondlengths depart from their bulk values, which are 2.35 \AA\ for
c-Si and 1.61 \AA\ for SiO$_2$. In addition, the angles of the subO Si
tetrahedra vary considerably: the Si--Si--Si angles vary from 99$^{\circ}$
to 125$^{\circ}$, the O--Si--Si angles vary from 96$^{\circ}$ to
126$^{\circ}$, while all O--Si--O angles remain around 106$^{\circ}$. It
is thus clear that the subO Si tetrahedra at the interfaces are distorted
as compared to bulk-Si tetrahedra.

\begin{figure}[htb]
\includegraphics*[width=3.0cm]{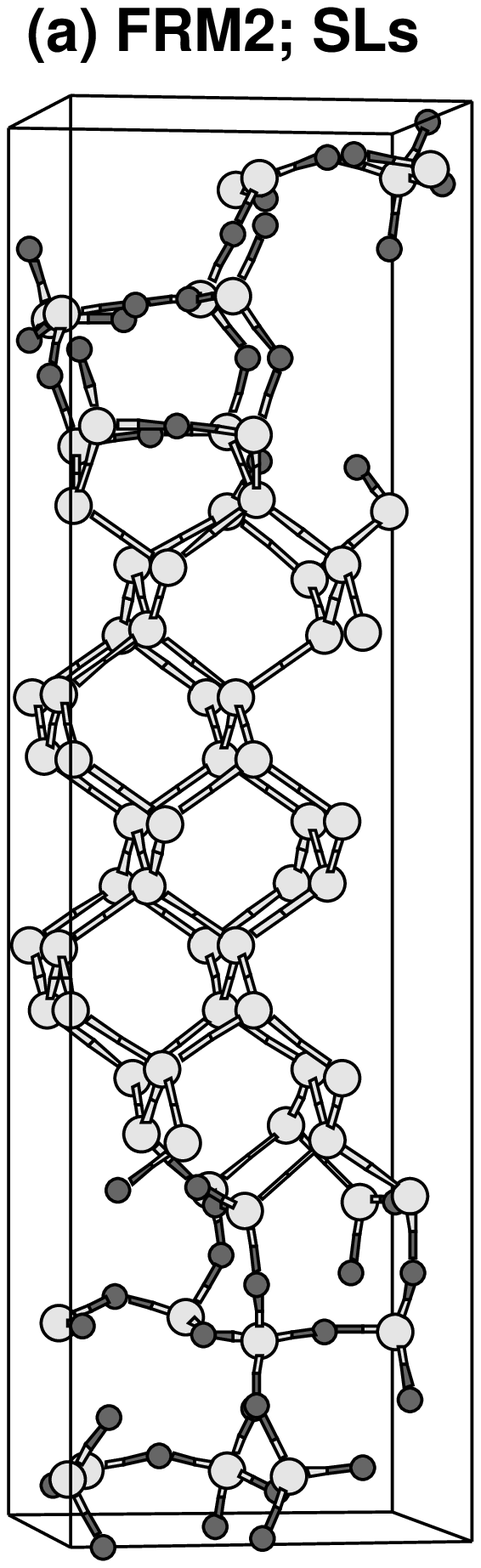}\includegraphics*[width=3.0cm]{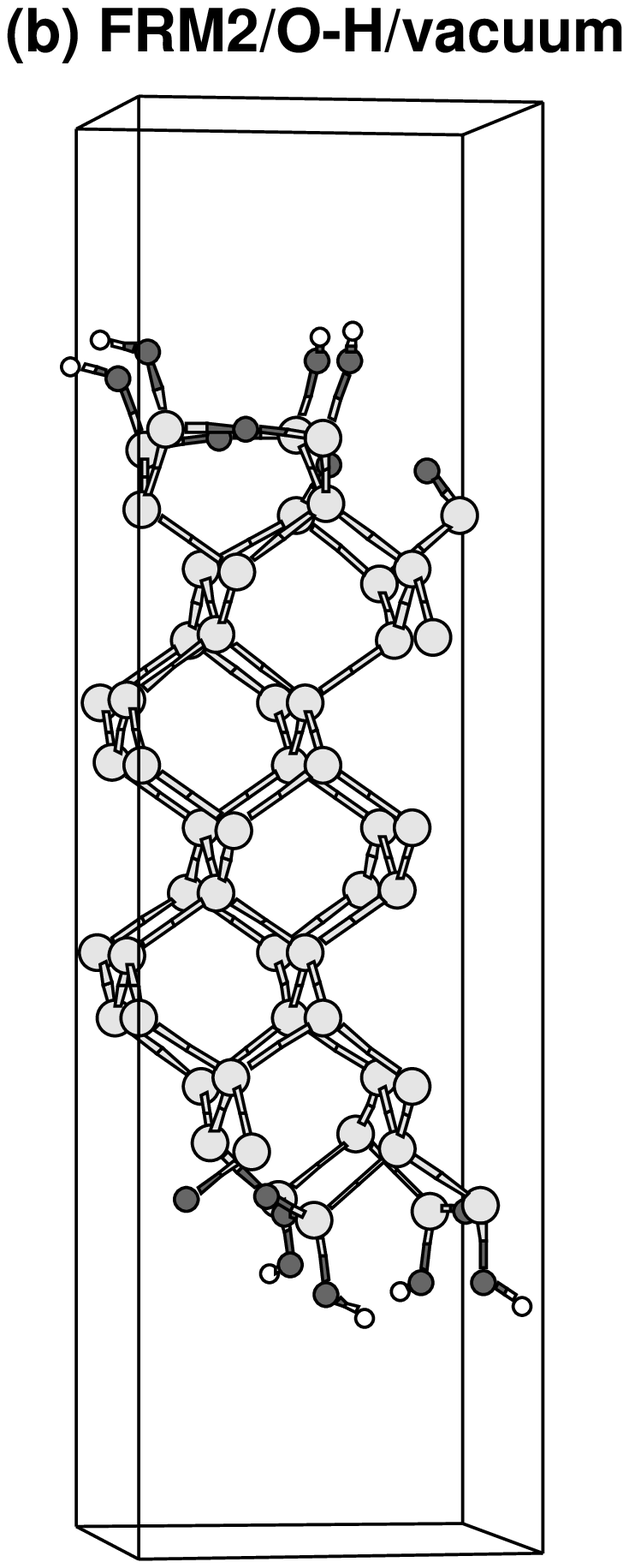}\includegraphics*[width=3.0cm]{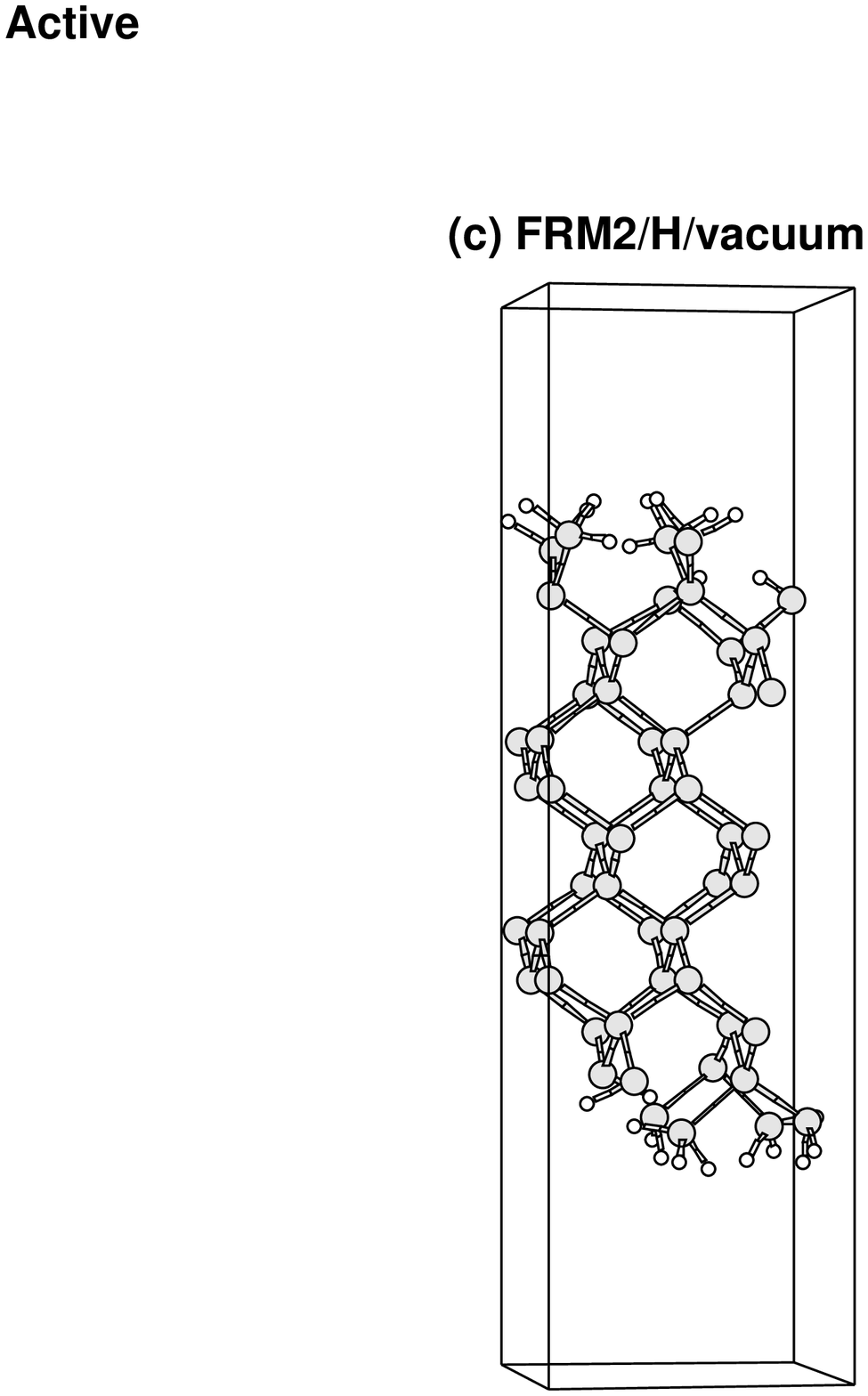}
  \caption{ (a) The FRM2 SLs (b) the FRM2/O-H/vacuum model
   (c) the FRM2/H/vacuum model.}
  \label{ConfinMod}
  \end{figure}

As discussed later on, the role of subO Si atoms was further studied using
the following variations of the FRM2:  First, we removed all Si$^{+4}$
atoms and attached the proper number of hydrogen atoms to neutralize the
excess charges. The H positions were then relaxed while keeping all the
silicon and oxygen atoms fixed. This structure thus contains Si$^0$ atoms
and Si$^{+n}$ subOs, where $n=1,2,3$ (i.e., $n \neq 4$).  A variation of
this structure was generated by removing all oxygen atoms and filling the
Si dangling bonds with H atoms, and again relaxing the H atoms. The final
Si-H bondlengths vary from 1.47 \AA\ to 1.53 \AA, after relaxation. This
final structure is thus subO-free and contains only Si$^{0}$ atoms, except
at the interface with the vacuum, where hydrogen atoms fill the dangling
bonds. These three confinement models are shown in Fig.~\ref{ConfinMod};
they will be referred to as ``FRM2'', ``FRM2/O-H/vacuum'' and
``FRM2/H/vacuum'', respectively.

\begin{figure}[htb]
   \includegraphics*[width=6cm]{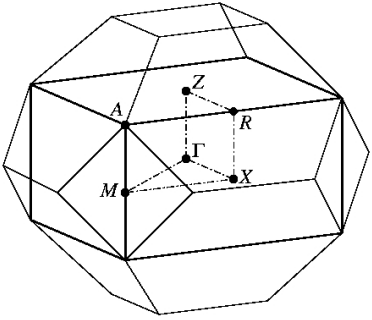}
  \caption{Definition of the SLs BZ superposed to the
  diamond-like BZ. The principal axis used for bandstructure
  calculation are also depicted.}
  \label{BZcSi}
  \end{figure}

In Fig.~\ref{BZcSi} we show the BZ of the supercell, the standard c-Si
diamond BZ, and the high symmetry axes used for the bandstructure
calculations. We also constructed the bulk c-Si structure in a supercell
of dimensions similar to that of the FRM SLs, so that comparisons can be
done within the same \textbf{k} space zone scheme.  This will be used in
the next three sections for comparisons of bandstructure as well as
absorption calculations.

\section{Quantum confinement}\label{confinement}

In this section, we discuss the nature of the confined states in the SLs.
We calculated the bandstructures of the three SL models --- FRM1, FRM2 and
FRM3 --- as well as the supplementary FRM2/O-H/vacuum structure, cf.\
Fig.~\ref{ConfinMod}(b). The latter is the ``ultimate'' in terms of
confinement, as the two interfaces with vacuum constitute infinite
potential walls. All bandstructures are analysed and compared within
equivalent supercell BZ. The growth axis of the SLs being the $z$-axis,
confinement effects are expected to take place in the \mbox{$X$--$R$} and
\mbox{$Z$--$\Gamma$} axis of the BZ (see Fig.~\ref{BZcSi} for axis
definitions).

The bandstructures of the three SLs and their total density of states
(DOS) are shown in Fig.~\ref{bands}. We find that the bandstructures in
the growth axis ($X$-$R$ and $Z$-$\Gamma$) are dispersionless, for all
models and all energies. In physical terms, dispersionless bandstructures
imply infinite effective masses, reflecting the strong confinement. The
DOS have a more abrupt variation in the valence bands than in the
conduction bands. However, DOS alone are not enough to fully understand
the optical processes involving the interband transitions, since the
weighting of the optical matrix elements is needed.  This is further
discussed in Section \ref{blueshift}.

\begin{figure}[htb]
\includegraphics*[width=4.2cm]{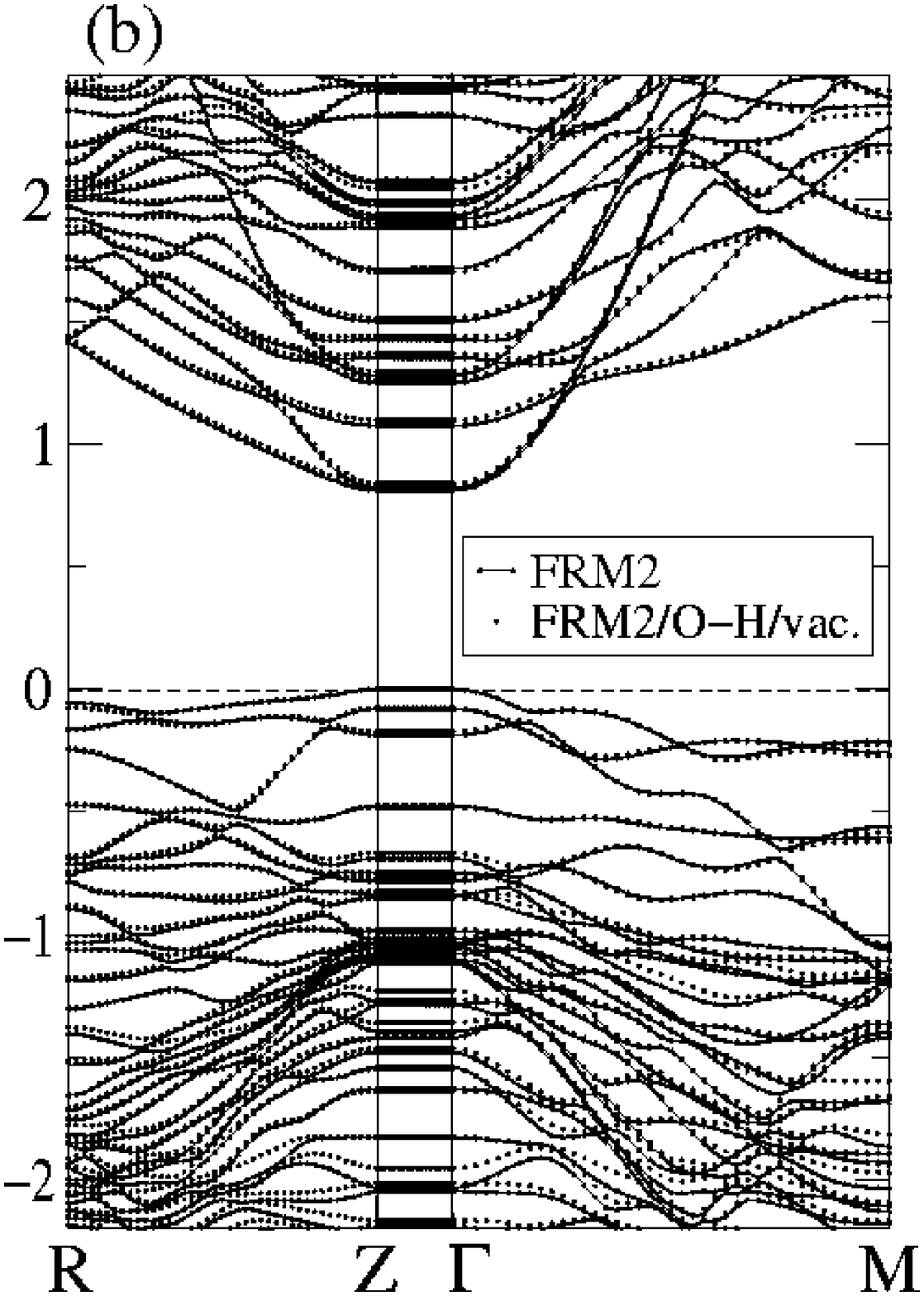}\includegraphics*[width=4.2cm]{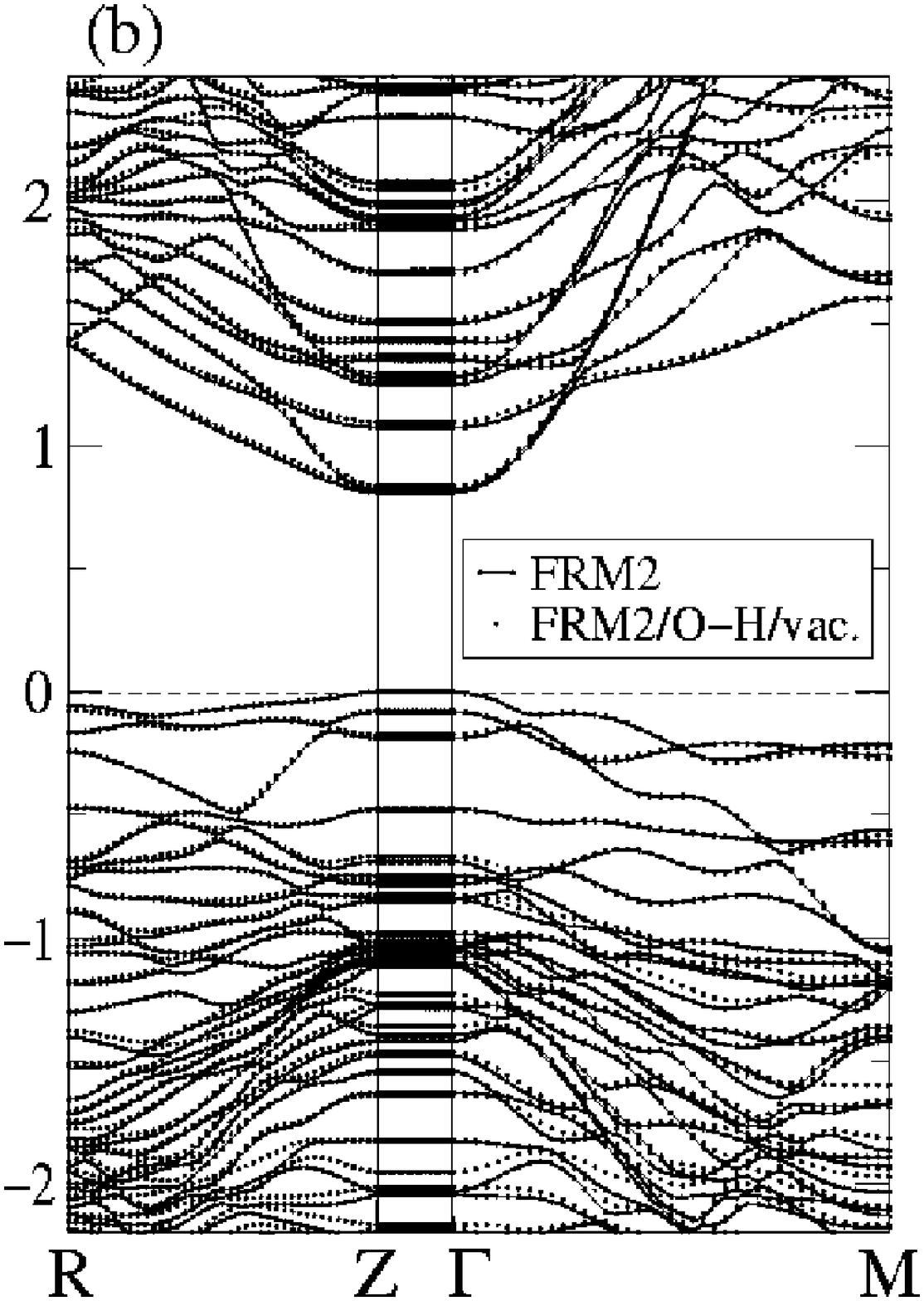}
  \caption{ Band structures
  of (a) c-Si in the (folded) SL BZ (b) comparison between the FRM2 SLs
  and the FRM2/O-H/vacuum bandstructures.}
  \label{bandSiOH}
  \end{figure}

Let us consider in more details one of the SL models, namely the FRM2. We
select the \mbox{$R$--$Z$--$\Gamma$--$M$} high-symmetry axis where the
major features, viz.\ the relevant energy gaps, appear.  The
bandstructures of the folded c-Si structure, the FRM2 SLs structure as
well as the FRM2/O-H/vacuum structure have been calculated and compared.
Figure \ref{bandSiOH}(a) and \ref{bandSiOH}(b) contain a synopsis of all
calculations for \textbf{k} points along \mbox{$R$--$Z$--$\Gamma$--$M$}.
Several conclusions can be drawn from these figures.

By comparing the bands for c-Si, Fig.~\ref{bandSiOH}(a), with those for
the FRM2 SL, Fig.~\ref{bandSiOH}(b) (solid lines), we see that folding
effects cannot by themselves explain the new band-structure. The bands in
the \mbox{$Z$--$\Gamma$} directions are totally modified by the
confinement. Although c-Si always has an optically indirect bandgap, this
bandgap becomes almost direct in its folded configuration, as can be seen
from Fig.~\ref{bandSiOH}(a), while for the SL the bandgap is
\emph{unequivocally} direct, and significantly increased. Moreover,
comparing the bands in a direction orthogonal to the growth axis, for
instance around $M$ in both Fig.~\ref{bandSiOH}(a) and
Fig.~\ref{bandSiOH}(b), we see that the valence bands are raised;  the
lowest conduction band in the \mbox{$\Gamma$--$M$} direction is pushed to
higher energies.  In addition, they exhibit less dispersion in the SL than
in c-Si, in general. Hence, confinement modifies the electronic properties
in the growth axis as well as in directions orthogonal to it.  This is
further analysed from optical absorption calculations, below.

We compare in Fig.~\ref{bandSiOH}(b) the bandstructures of the FRM2 and
FRM2/O-H/vacuum models. The positions of the Si$^0$ atoms, as well as the
subO Si in the two structures, are identical. The solid lines displays the
bandstructures of the FRM2, while the dots display the bands of the
FRM2/O-H/vacuum structure. It is clear that the two bandstructures are
nearly identical.  This calculation shows that the SiO$_2$ layers act as
virtually impenetrable barriers. The electronic properties of hypothetical
SL structures having only subO Si -- i.e., no Si$^{+4}$ of SiO$_2$ -- and
positioned at the subO Si sites, would give nearly identical electronic
properties as Si/SiO$_2$ SLs, for energies close to the band gap. Our
calculations show that the electronic wavefunctions die out at the
suboxide Si atoms of the interfaces. Thus the barrier is sharply located
just behind the subO Si atoms, and therefore just two atomic layers could
be used as a barrier without altering the electronic properties, when
energies involved (in the device) remain close to the energy gap.  The
influence of the interface subO Si atoms is further analysed next.

\section{Role of interfaces}\label{interface}

\begin{figure}[htb]
  \includegraphics*[width=8.2cm]{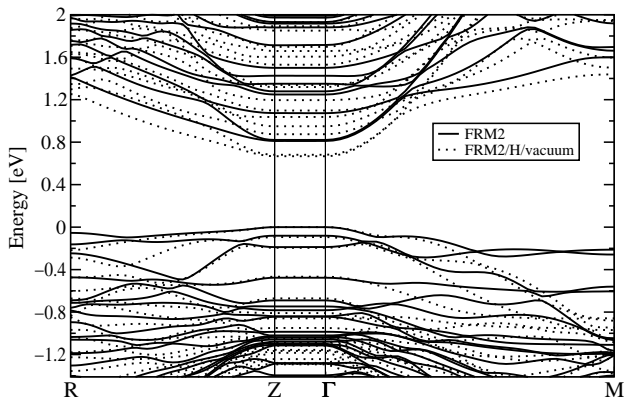}
  \caption{ Comparison of the bandstructures for
  FRM2/H/vacuum and FRM2 SLs. }
  \label{bandH}
  \end{figure}

In order to assess the role of the subO ions on the electronic properties,
we calculated the bandstructures of the FRM2/H/vacuum structure,
Fig.~\ref{ConfinMod}(c), which contains {\it no} subO Si; the dangling
bonds have been filled by hydrogen atoms and hydrogen atoms (only) have
been structurally relaxed.  Figure \ref{bandH} summarizes the results. The
LDA band gap is still direct but significantly lowered, from 0.81 eV in
the FRM2 to 0.67 eV in the FRM2/H/vacuum structure. Interface
reconstruction (as discussed in section \ref{Ionic} and
Fig.~\ref{couches}) thus has significant impact on the electronic
properties. The bands in the plane orthogonal to the growth axis are quite
different from FRM2; e.g., the valence band is lowered near the $M$ and
$R$ points, becoming similar to the c-Si bandstructure. We thus conclude
that the subO Si atoms have two effects: (i) increase the bandgap and (ii)
produce dispersionless valence bands. The charge states in the subO are
responsible for the increase in the bandgap, while the strongly increased
valence-band offset leads to essentially dispersionless bands.

\section{Blueshift and optical enhancement}\label{blueshift}

The matrix elements entering the calculations of the optical properties
are often approximated as a constant in a given range of energies.
However, such an approximation is inadequate for elucidating the enhanced
luminescence in Si/SiO$_2$ SLs. This section deals with calculating the
absorption coefficient within the Fermi Golden rule and
interband-transition theory.

Since the Si-layer thickness $L_{\rm Si}$ is relevant to the energy-shift
in SLs, this quantity needs first to be defined.  This involves some
uncertainty associated with the interface thicknesses $L_{\rm subO}$.

The interface thickness $L_{\rm subO}$ was estimated to be $\sim$1.60 \AA\
from our calculations of the subO Si region. This is the largest distance
(projected onto the $z$ axis) between any two subO Si atoms. Hence the
upper-bound to $L_{\rm Si}$ are, 11.17 \AA, 16.58 \AA\ and 22.01 \AA\ for
the FRM1, FRM2 and FRM3 respectively, while the lower-bound thicknesses
are simply $L_{\rm Si}-2L_{\rm subO}$. We define the Si thickness in the
SLs to include the interface subO Si atoms as well (corresponding to the
upper bound). This choice is made since subO Si (Si$^{+1}$, Si$^{+2}$ or
Si$^{+3}$ ) atoms contribute to the electronic properties, as do bulk-Si
atoms (Si$^0$ atoms); for instance, we showed above [see e.g.  
Fig.~\ref{bandSiOH}(b)] that the bandstructures of the FRM2 SLs and the
FRM2/O-H/vacuum systems overlap, and indicated where the effective barrier
begins.

\begin{figure}[htb]
   \includegraphics*[width=8.2cm]{bandFRM1.eps}
   \includegraphics*[width=8.2cm]{bandFRM2.eps}
   \includegraphics*[width=8.2cm]{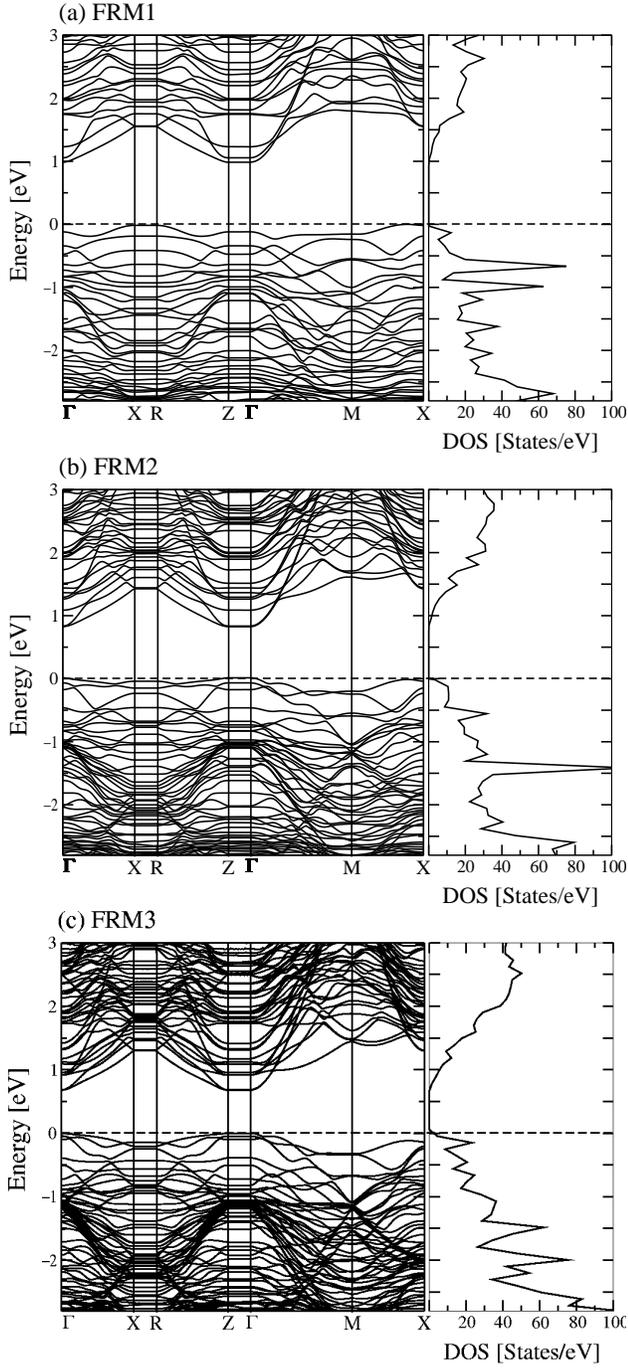}
  \caption{  Band structures and density of states (DOS)
            of the three SL models. The DOS are 
            calculated using 
            \mbox{(7 $\times$ 7 $\times$ 1)} \textbf{k}-point grid,
            which corresponds to 144 irreducible tetrahedra.} 
  \label{bands}
  \end{figure}

The band gaps of the FRM1, FRM2 and FRM3 (see the bandstructures in
Fig.~\ref{bands}) are direct except for the FRM1 where the band gap is
\emph{nearly} direct with only 0.12 eV between the direct and indirect
transitions.  The values of the gap are \mbox{0.99 eV}, \mbox{0.81 eV} and
\mbox{0.68 eV} for the FRM1, FRM2 and FRM3, respectively. The direct
transition at $\Gamma$ for the FRM1 equals \mbox{1.11 eV}. For the FRM2
and the FRM3, the band gaps are direct and located on the whole
$Z$--$\Gamma$ axis [see Fig.~\ref{bands}(b) and (c)]. Direct transitions
can thus be achieved between the valence band and the conduction band,
along the $Z-\Gamma$ line of the BZ.  We thus obtain, under the LDA, a
blue shift
 $$
 \mbox{0.68 eV} \longrightarrow \mbox{0.81 eV} \longrightarrow
 \mbox{0.99 eV}
 $$
 with increased confinement
 $$
 \mbox{2.2 nm} \longrightarrow \mbox{1.7 nm} \longrightarrow
 \mbox{1.1 nm}.
 $$
 However, these values for the energy gaps are much lower than the
experimental ones. It is well known that DFT within the LDA underestimates
the energy gaps of semiconductors and insulators. For c-Si, the DFT-LDA
gap is approximately 0.6 eV less than the experimental value. For the
$\beta$-cristobalite phase of SiO$_2$, in the group $I\bar{4}2d$, the
DFT-LDA energy gap is \mbox{5.8 eV} while the experimental value is about
3 eV higher, i.e., \mbox{$\sim$9 eV}.  Approximate, but realistic, band
gaps can be obtained by adding 0.6 eV overall:
 $$
 \begin{array}{ccccc}
 \mbox{\bf 1.28 eV} & \longrightarrow & \mbox{\bf 1.41 eV} &
 \longrightarrow & \mbox{\bf 1.59 eV}. \\
 \mbox{(2.2 nm)} & & \mbox{(1.7 nm)} & & \mbox{(1.1 nm)}
 \end{array}
 $$
 These energy gaps correspond to the lower bound of the
experiments\cite{Novikov} and lie in the visible spectrum. However, these
gaps are still somewhat lower than the experimental
values.\cite{LuLockBar,Kanemitsu,Khriachtchev,Mulloni,Kanemitsu_bis} The
discrepancy can be explained by recrystallization processes, which lead to
the formation of nanoclusters that would increase the confinement, and
correspondingly the measured energy gaps.\cite{ZachariasMRS} The analysis
of such a behaviour, which is beyond the scope of the present work, would
require zero-dimensional-confined model structures.

The absorption of the three SL models and c-Si have been calculated both
in the diamond-like BZ and in the SLs BZ.  For c-Si in the diamond-like
BZ, we used \mbox{(20 $\times$ 20 $\times$ 20)} \textbf{k}
points,\cite{Adolph} while in the SLs BZ, \mbox{(7 $\times$ 7 $\times$ 2)}
\textbf{k} points are used. Calculations using more \textbf{k} points,
viz.\ \mbox{(8 $\times$ 8 $\times$ 3)}, show that \mbox{(7 $\times$ 7
$\times$ 2)} is quite sufficient to recover the form of the absorption
curve for all the SL models. The purpose of calculating the absorption of
c-Si in two different BZ is to estimate errors, first due to zone folding
effects (which cause round-off errors, leading to non-absolutely-null
transitions at the onset)\cite{Furthmuller} and, second, to the
tetrahedron method itself which needs large amounts of \textbf{k} points.  
The broadening in the absorption curves has been fixed to \mbox{0.015 eV}
for all the absorption curves discussed below, as suggested by
Fuggle.\cite{Fuggle}

\begin{figure}[htb]
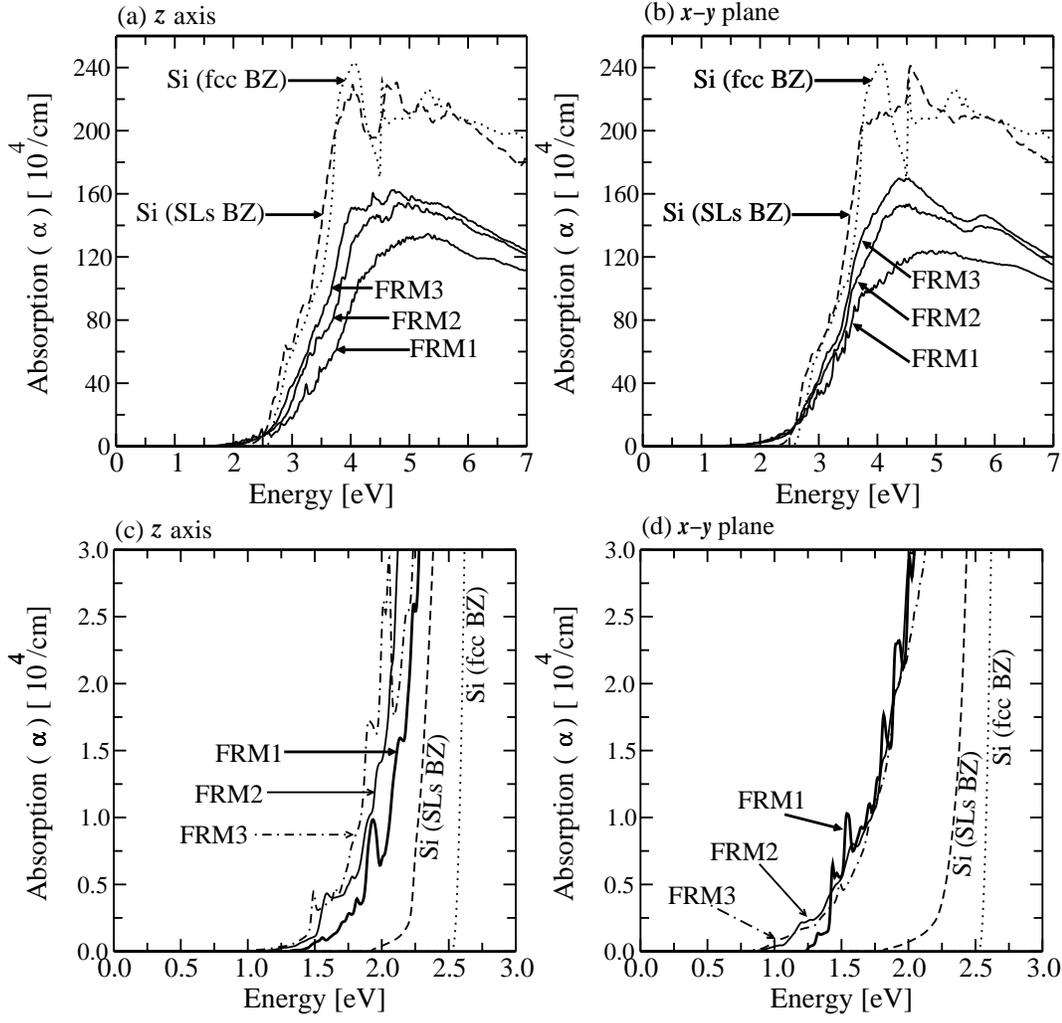

\includegraphics*[width=7cm]{absZtot.eps}\includegraphics*[width=7cm]{absXYtot.eps}
\includegraphics*[width=7cm]{absZzoom.eps}\includegraphics*[width=7cm]{absXYzoom.eps}
  \caption{ Absorption coefficient of the SL models as compared to c-Si.
    (a) and (c) are the absorption in the growth axis; (b) and (d)
    are the absorption in the plane orthogonal to the growth axis. }
  \label{Absorp}
  \end{figure}

Figure \ref{Absorp} shows the absorption results. Panels (a) and (b) give
an overall view of the absorption curves for the $z$ axis in (a) and the
$x$-$y$ plane in (b). Panels (c) and (d) show the absorption at the onset,
for the $z$ axis in (c) and the $x$-$y$ plane in (d).  In all cases, we
included the absorption of c-Si calculated in the diamond-like BZ, as well
as the one in the SLs BZ. Direct comparison of the two c-Si absorption
curves give an estimate of imprecisions due to zone foldings and
intermediate number of \textbf{k} point effects. It shows that the
absorption is slightly underestimated in the SL calculations; e.g., in
Fig.~\ref{Absorp}(c), the onset of absorption of c-Si in the SLs BZ take
place at 2.0 eV while in the diamond-like BZ the onset happens at the
correct value of 2.52 eV (which is for c-Si the LDA direct transition at
$\Gamma$).  This numerical effect cannot be avoided and will arise, as
well, in any Si/SiO$_2$ supercell. Hence, all comparison of the SLs
absorption must be made with c-Si calculated in the equivalent SLs BZ,
i.e., within equivalent {\bf k} space zone schemes.

Since the SLs are fabricated with the objective of changing the indirect
gap to a direct gap, we now discuss the absorption threshold region.
Comparison at the onset of absorption from Fig.~\ref{Absorp}(c) or (d)
shows that all absorption curves have a lower energy threshold than
\emph{both} the c-Si absorption curves, and especially below the one
calculated in equivalent SLs BZ having equal number of \textbf{k} points.
That is, {\it the SLs show absorption (and emission) in the spectral
region above the indirect gap of c-Si and below the direct gap of c-Si}.
This shows that the absorption in \emph{all} confined Si/SiO$_2$ SL models
is enhanced, compared to \mbox{c-Si}; the transitions are direct in SLs
and have an active oscillator strength. For the folded c-Si energy bands,
the lower bands above the Fermi level, and the corresponding oscillator
strength, remain dark; in other words, the optical matrix elements of the
SL BZ of c-Si are null up to $\sim$2.0 eV. This result clearly
demonstrates the enhancement of the absorption (and emission) mechanisms
in confined Si structures.  Furthermore, upon inspection of the absorption
curves in the plane orthogonal to the growth axis [Fig.~\ref{Absorp}(d)],
we note that the energy thresholds of the absorption are all below c-Si:
thus, the $x$-$y$ absorption of the FRMs takes place approximately at
their respective direct energy gaps, and then behave in a similar manner,
as expected from the similarity of the SLs bandstructures, in this plane.

We examine, finally, the higher-energy region which corresponds to the
usual direct transition (3 - 6 eV) in c-Si.  Even here,
Fig.~\ref{Absorp}(a) demonstrates a blueshift with increased confinement
in the $z$ axis. The overall absorption maxima for FRM1-FRM3 are at 5.28
eV, 4.83 eV, and 4.71 eV, with intensities of 136, 155 and 162 ($\times
10^4$/cm) respectively.  For c-Si in the SLs BZ (to ensure comparable
precision in the calculations) the second peak, i.e., the maximum, take
place at 4.70 eV (with absorption equal to \mbox{231 $\times$ 10$^4$/cm}),
while the first peak is at 4.0 eV (and with absorption equal to \mbox{229
$\times 10^4$/cm}). We emphazise that there is still a slight blueshift in
the $x$-$y$ plane orthogonal to the growth axis, but less pronounced than
in the growth axis.

\section{Concluding remarks}

In this work, the structural, electronic and optical properties of
Si/SiO$_2$ superlattices have been studied on the basis of
structurally-relaxed models. These SL models, which contain no hydrogen
atoms at the Si/SiO$_2$ interfaces, exhibit enhanced optical
absorption/emission, as observed in experiment; this can be attributed to
the presence of silicon dioxide barriers. In experiments performed under
ultra-high vacuum conditions, the oxidization process would predominantly
give rise to oxide bonds at the interfaces, but still, few hydrogen atoms
are expected to be present and fill some of the remaining dangling bonds.

Our calculations show that the oxide barriers are central to the optical
enhancement in SLs.  It is well known that hydrogen atoms play a similar
role in amorphous silicon by filling dangling bonds. This suggests that it
might also be the case in Si/SiO$_2$ SLs, where hydrogen atoms fill extra
dangling bonds, and thus would amplify the optical enhancement effect
already exerted by the oxide barriers. Further studies are needed to
ascertain this.

We have shown that suboxide Si atoms at the interfaces act as virtually
impenetrable barriers. The active barrier thickness thus corresponds to
the suboxide Si layer --- only 1.6 \AA\ in our models. The confined Si
layer thus consists of bulk Si \emph{and} suboxide Si atoms. Suboxide Si
atoms at the interfaces modify the electronic properties in two manners:
they (i) increase the energy gap and (ii) lead to dispersionless band
structures, which increases the transition probabilities.

Other confinement models (zero-dimensional structures) and interface
effects will be considered in future work. For instance, inclusion of
other atomic species --- such as nitrogen that would generate subnitric Si
atoms at the Si/SiO$_2$ interface --- are expected to modify the
electronic properties, and hence enhance the optical absorption/emission
spectra.

\vspace{0.5cm}

{\it Acknowledgments} -- It is a pleasure to thank Dr.\ Jurgen
Furthm\"uller for help with the optical calculations in VASP.  We
gratefully acknowledge Dr.\ Zheng-Hong Lu for helpful discussions and
suggestions. This work is supported by grants from the Natural Sciences
and Engineering Research Council (NSERC) of Canada and the ``Fonds pour la
formation de chercheurs et l'aide {\`a} la recherche'' (FCAR) of the
Province of Qu{\'e}bec. We are indebted to the ``R\'eseau qu\'eb\'ecois de
calcul de haute performance'' (RQCHP) for generous allocations of computer
resources.

\newpage

\begin{references}
 \bibitem{LuLockBar} Z.H. Lu, D.J. Lockwood, and J.-M. Baribeau, Nature
    \textbf{ 378}, 258 (1995).
 \bibitem{Kanemitsu} Y. Kanemitsu and S. Okamoto, Phys. Rev. B
    \textbf{56}, R15561 (1997).
 \bibitem{Novikov} S.V. Novikov, J. Sinkkonen, O. Kilpel\"a, and S. V.
    Gastev, J. Vac. Sci. Technol. B \textbf{ 15}, 1471 (1997).
 \bibitem{Khriachtchev} L. Khriachtchev, M. R\"as\"anen, S. Novikov,
    O. Kilpel\"a, and J. Sinkkonen, J. Appl. Phys \textbf{ 86}, 5601
    (1999).
 \bibitem{Mulloni} V. Mulloni, R. Chierchia, C. Mazzeleni, G. Pucker, L.
    Pavesi, and P. Bellutti, Philos. Mag. B \textbf{ 80}, 705 (2000).
 \bibitem{Kanemitsu_bis} Y. Kanemitsu, M. Liboshi, and T. Kushida, Appl.
    Phys. Lett. \textbf{ 76}, 2200 (2000).
 \bibitem{Canham} L.T. Canham, Appl. Phys. Lett. \textbf{ 57}, 1046
    (1990).
 \bibitem{LehmanGosele} V. Lehman and U. G\"osele, Appl. Phys. Lett.
    \textbf{58}, 856 (1991).
 \bibitem{Tsybeskov} L. Tsyberkov, K.L. Moore, D.G. Hall, and P.M.
    Fauchet, Phys. Rev. B \textbf{ 54}, R8361 (1996).
 \bibitem{Cheylan} S. Cheylan and R.G. Elliman, Appl. Phys. Lett.
    \textbf{78}, 1912 (2001).
 \bibitem{Ng} W.L. Ng, M.A. Louren\c{c}o, R.M. Gwilliam, S. Ledain, G.
    Shao, and K.P. Homewood, Nature \textbf{ 410}, 192 (2001).
 \bibitem{ericson} P. Schmuki, L.E.  Erickson, and D.J. Lockwood,
    Phys. Rev. Lett. {\bf 80}, 4060 (1998).
 \bibitem{porous} S. Schuppler, S.L. Friedman, M.A. Marcus, D.L. Adler,
    Y.-H. Xie, F.M. Ross, Y.J. Chabal, T.D. Harris, L.E. Brus, W.L.
    Brown, E.E. Chaban, P.F. Szajowski, S.B. Christman, and P.H. Citrin,
    Phys. Rev. B {\bf 52}, 4910 (1995); D.J. Lockwood, A. Wang, and
    B. Bryskiewicz, Solid State Comm. {\bf 89}, 587 (1994).
 \bibitem{Luprivate} Z.H. Lu (private communication).
 \bibitem{Zacharias} M. Zacharias, J. Bl\"asing, K. Hirschman, L.
   Tsybeskov, P.M. Fauchet, J. Non-Crys. Solids {\bf 266-269}, 640 (2000).
 \bibitem{ZachariasAPL} M. Zacharias, J. Bl\"{a}sing,
    P. Veit, L. Tsybeskov, K. Hirschman, and P. M. Fauchet, Appl. Phys.
    Lett. \textbf{74}, 2614 (1999).
 \bibitem{Stirling} A. Stirling, A. Pasquarello, J. -C. Charlier, R. Car,
   Phys. Rev. Lett. \textbf{85}, 2773 (2000).
 \bibitem{Delley} B. Delley and E.F. Steigmeier, Appl. Phys. Lett.
    {\bf 67}, 2370 (1995).
 \bibitem{Kageshima} H. Kageshima and K. Shiraishi, in \emph{Materials and
    Devices for Silicon-Based Optoelectronics}, edited by J.E. Cunningham,
    S. Coffa, A. Polman, and R. Soref, Mater. Res. Soc. Symp.
    Proc. No.\textbf{ 486} (Material Research Society, Pittsburgh, 1998),
    p. 337.
 \bibitem{Pukkinen} M.P.J. Pukkinen, T. Korhonen, K. Kokko, and I.J.
    V\"ayrynen, Phys. Stat. Sol. {\bf 214}, R17 (1999).
 \bibitem{Degoli} E., Degoli, S. Ossicini, Surf. Sci. {\bf 470}, 32
   (2000).
 \bibitem{Carrier} P. Carrier, L.J. Lewis, and M.W.C. Dharma-wardana,
    Phys. Rev. B. {\bf 64}, 195330-1 (2001).
 \bibitem{Agrawal} B.K. Agrawal and S. Agrawal, Appl. Phys. Lett.
    {\bf 77}, 3039 (2000).
 \bibitem{Fuggle} J.C. Fuggle, in \textit{Unoccupied Electronic States},
    edited by J.C. Fuggle and J.E. Inglesfield,
    Topics in Applied Physics Vol.\ \textbf{69}
    (Springer-Verlag, Berlin, 1992).
 \bibitem{SiegerLuHimpsel}
    M.T. Sieger, D.A. Luh, T. Miller, and T.-C. Chiang,
    Phys. Rev. Lett. {\bf 77}, 2758 (1996);
    Z.H. Lu, M.J. Graham, D.T. Jiang, and K.H. Tan,
    Appl. Phys. Lett. {\bf 63}, 2941 (1993);
    F.J. Himpsel, F.R. McFeely, A. Taleb-Ibrahimi,
    J.A. Yarmoff, and G. Holliger, Phys. Rev. B {\bf 38}, 6084 (1988).
 \bibitem{Lockwood} D.J. Lockwood, Z.H. Lu, and J.-M. Baribeau, Phys. Rev.
    Lett. {\bf 76}, 539 (1996).
 \bibitem{Pasquarello} A. Pasquarello, M.S. Hybertsen, and R. Car, Appl.
    Surf. Sci. \textbf{104/105}, 317 (1996);
    A. Pasquarello, M.S. Hybertsen, and R. Car, Appl. Phys. Lett.
    {\bf 68}, 625 (1996).
 \bibitem{Neaton} J.B. Neaton, D.A. Muller, and N.W. Ashcroft, Phys. Rev.
    Lett. {\bf 85}, 1298 (2000).
 \bibitem{Tu} Y. Tu and J. Tersoff, Phys. Rev. Lett. {\bf 84}, 4393
    (2000).
 \bibitem{NgInterface} K.-O. Ng and D. Vanderbilt, Phys. Rev. B {\bf 59},
    10132 (1999).
 \bibitem{KageshimaInterface} H. Kageshima and K. Shiraishi,
    Phys. Rev. Lett. {\bf 81}, 5936 (1998).
 \bibitem{HermanBatra} F. Herman and I.P. Batra, in {\it The Physics of
    SiO$_2$ and its interfaces} edited by S.T. Pantelides (Pergamon,
    Oxford, 1978).
 \bibitem{TitDharmaDBMBOM} N. Tit and M.W.C. Dharma-wardana, Physics
    Letters A \textbf{ 254}, 233 (1999).
 \bibitem{TitDharma} N. Tit and M.W.C. Dharma-wardana, J. Appl. Phys.
    \textbf{ 86}, 1 (1999).
 \bibitem{tran} M. Tran, N. Tit, and M.W.C. Dharma-wardana, Appl. Phys.
    Lett. {\bf 75}, 4136 (1999).
 \bibitem{VASPref} G. Kresse and J. Furthm\"uller, computer code
    \textbf{VASP 4.4} (Vienna University of Technology, Vienna, 1997)
    [Improved and updated Unix version of
    the original copyrighted VASP/VAMP code, which was published by G.
    Kresse and J. Furthm\"uller, Comput. Mater. Sci. {\bf 6}, 15 (1996)].
 \bibitem{KresseJoubert} G. Kresse and D. Joubert, Phys. Rev. B {\bf 59},
    1758 (1999).
 \bibitem{HohenbergKohn} P. Hohenberg and W. Kohn, Phys. Rev. {\bf 136},
    B864 (1964).
 \bibitem{KohnSham} W. Kohn and L.J. Sham, Phys. Rev. {\bf 140}, A1133
    (1965).
 \bibitem{Payne} M.C. Payne, M.P. Teter, D.C. Allan, T.A. Arias, and
    J.D. Joannopoulos, Rev. Mod. Phys. {\bf 64}, 1045 (1992).
 \bibitem{BlochlPAW} P.E. Bl\"ochl, Phys. Rev. B {\bf 50}, 17953 (1994).
 \bibitem{Singh} D.J. Singh, \textit{Planewaves, pseudopotentials and the
    LAPW method}, (Kluwer Academic Publishers, Norwell, 1994).
 \bibitem{Adolph} B. Adolph, J. Furthm\"uller, and F. Bechstedt, Phys.
    Rev. B {\bf 63}, 125108-1 (2001).
 \bibitem{Cardona} P.Y. Yu and M. Cardona, \emph{Fundamentals of
    Semiconductors} (Springer, New-York, 1996);
    G. Baym, \emph{Lectures on Quantum
    mechanics}, (Addison-Wesley, Reading, 1993); L. Ley in \emph{The
    Physics of Hydrogenated Amorphous Silicon II}, edited by J.D.
    Joannopoulos and G. Lucovsky, Topics in Applied Physics, Vol.\
    {\bf 56} (Springer-Verlag, Berlin, 1984).
 \bibitem{BlochlTET} P.E. Bl\"ochl, Phys. Rev. B {\bf 49}, 16223 (1994).
 \bibitem{Chang} E.K. Chang, M. Rohlfing, and S.G. Louie, Phys. Rev. Lett.
    {\bf 85}, 2613 (2000); S. Albrecht, L. Reining, R. Del Sole, and G.
    Onida, Phys. Rev. Lett. {\bf 80}, 4510 (1998).
 \bibitem{ZachariasMRS} M. Zacharias, J. Heitmann, and U. G\"{o}sele,
 MRS Bulletin \textbf{26}, 975 (2001).
 \bibitem{Furthmuller} J. Furthm\"uller (private communication).
 \end{references}
\end{document}